# Quality Control and Validation Boundaries in a Triple Helix of University-Industry-Government: 'Mode 2' and the Future of University Research

Yuko Fujigaki

Department of International and Interdisciplinary Studies, University of Tokyo,

3-8-1 Komaba, Megurao-ku Tokyo 153-8902, Japan

FAX: +81 3 5454 6990 e-mail: fujigaki@idea.c.u-tokyo.ac.jp)

and

Loet Leydesdorff

Science & Technology Dynamics, Faculty of Social and Behavioural Sciences,

OZ Achterburgwal 237, 1012 DL  Amsterdam, The Netherlands

Tel.: +31 20  525 65 98, Fax : +31 20  525 2086, e mail: loet@leydesdorff.net;

http://www.leydesdorff.net/

Introduction

Validation Boundaries as a Conceptual Tool

Quality Control for Mode 2 Research

  *Market force*

  *Problem-solving in public sphere*

  *Synthesis -- Linking Validation Boundaries*

Institutional Boundaries and Validation Boundaries

  *Implication for Comparative Studies*

The Future of the Academe

**Abstract**

How is quality control organized in the new "Mode 2" of the production of scientific knowledge?  When institutional boundaries are increasingly blurred in a Triple Helix of University-Industry-Government relations, criteria for quality control in the production of scientific knowledge can be expected to change at the interfaces. The categorization in terms of two modes of knowledge production was introduced by Gibbons et al. (1994) in order to describe changes in the networks of scientific communications (funding patterns, research configurations, styles of knowledge management, etc.). These changes were mainly specified as institutional parameters in order to deal with the subjects of R&D management and S&T policies, that is, ex ante (Spiegel-R ing 1973; Van den Daele et al. 1979). We focus on the "validation boundaries" emerging from the differences between Mode 1 and Mode 2; that is, on the criteria for quality control that can analytically and reflexively be brought to the fore ex post. The shift from an institutional frame of reference to a focus on the dynamics of communications enables us to clarify several problems in the discussion of
the future of university research.

**Introduction**

The difference between "Mode 1" and "Mode 2" (Gibbons *et al.* 1994) has sometimes been considered as a reappraisal of the old distinction between basic science and the contexts of its application (e.g., Weingart 1997; Godin 1998). Does Mode 2 differ from Mode 1 epistemologically or is the difference only contextual? How could it be possible that a different context would change the validity of a knowledge claim?

In our opinion, various validation mechanisms have always interacted in scientific developments; Haynes (1990), for example, noted that the *Journal of Internal Medicine* has often accepted papers which use questionable scientific methodologies, such as uncontrolled trials. In his opinion, such methodologically problematic papers—being in their early stages and therefore not having undergone rigorous testing—are essential for communication among scientists, whereas more rigorous testing and strictly controlled studies are needed for communication among between clinicians. Since clinicians encounter the public in their medical practice, one witnesses a "reverse tendency" between "Mode 2" and "Mode 1" research. In this case, the use of research in contexts of application requires a more thorough specification of the substantive issues and the methodological standards are also higher.

The epistemological codes of what is considered a valuable contribution can differ between communication within a scientific community and communication with an external audience. Furthermore, the distinction between scientific excellence and practical relevance is analytically independent of the rule of the market (Lundvall, 1988). Thus, the meaning of what Gibbons *et al.* (1994) called "contexts of application" has to be specified more precisely. Several rules or principles may operate can be operating in parallel given a pluriform society. The sciences compete not only in terms of their effectiveness on relevant markets, but also in terms of other user-criteria such as quality of life and sustainability. For example, one can analyze the value of "problem solving" in the public sphere. The value of an insight for public decision-making may be completely different from its value on the market.

*Validation boundaries*

The concept of boundaries enables us to understand the issue of quality criteria of different audience. Boundaries within and between bodies of knowledge can be validated, legitimated, and utilized by communities of users. These users may have a different access to the substantive content of the knowledge being used. We propose to distinguish between institutional boundaries and validation boundaries: "validation boundaries" (Fujigaki 1998) in bodies of knowledge remain products of communication. They can therefore be more flexible than institutional boundaries.

The difference between validation boundaries and institutional boundaries provides the communication with different meanings and thus makes it possible for researchers and users to "translate" between domains. The standards developed in these translations can have critical functions in reshaping the institutions (Etzkowitz and Leydesdorff 1997). The analysis of the formulation and interaction of validation boundaries in relation to the boundaries of institutional units can thus be useful when the future development of scientific knowledge is discussed in terms of scientific communications. One is able to raise the question of how quality control can be "self-organized" by the communication (Leydesdorff 1995).

From this (neo-) evolutionary perspective (Maturana & Varela 1980; Luhmann 1990), one would *not* expect the further codification and integration of scientific communications in two contexts to lead to "dedifferentiation". Integration is then achieved by adding a reflexive next-order level. The translations between communications using different codes can enrich the system with next-layers of potentially higher-order codification. However, this additional layer cannot be stable, although it may induce and reinforce institutional (re-)organization over time.

The next-order level tends to remain an interface under construction since the emerging order is embedded in the communications on which it rests. The perspective enables us to consider qality control in Mode 1 and quality control in Mode 2 research as categories that can be different in terms of the system of reference and the levels of integration. The two levels can then no longer be summarized under a single – that is, one-dimensional – concept of quality control. Quality is no longer a supra-historical category, but one has to specify a system of reference.

The new models for the production of scientific knowledge in networks of relations between institutional units –e.g., university-industry-government relations— raise the question of how "quality control" of scientific communications can be organized given the fluidity of these network relations. How can one differentiate and integrate the knowledge production in contexts of application and in disciplinary contexts without compromising? How is one able to balance and/or trade-off between the two?

**Validation Boundaries as a Conceptual Tool**

Advanced scientific knowledge production ("Mode 1") usually results in the submission of publications to scientific journals. The quality control of these products of scientific knowledge is achieved by the review system used by the journals. The legitimacy and validity of knowledge is controlled and reinforced through the process of judging whether or not submitted papers can be accepted. Some papers are accepted and others are rejected and this accepted-rejected-action recursively constructs the validation-boundary of knowledge production (Fujigaki, 1998).[1] A validation boundary remains (partly) "invisible" to the actors involved (Crane, 1972), and this latent-operation of acceptance and rejection can be considered one of the essential functions of peer review (Fuller 1998a).[2]

This specific form of quality control, however, can be maintained by the scientific community (quasi-) autonomously because it has functions for the development of the system itself. In our opinion, it is a property of the specific form of communication in science, which should not be confused with actor-categories. Whereas, for example, the concept of trans-epistemic arenas was specified in order to describe and explain specific actions observable in laboratories (cf. Knorr-Cetina, 1982), the concept of a validation boundary can be used for discussing quality control at the field level. In empirical science studies research one can deal with this process of quality control by observing the results of the acceptance or the rejection of different papers.

Note that validation boundaries exist only operationally, while disciplinary boundaries (e.g., Fuller, 1988) are defined structurally and are therefore institutionally observable. Although there are several studies which deconstruct peer review systems as questionable processes in empirical research (e.g., Franz, 1974; Horrobin, 1990; Paker, 1997), their epistemological status as processes of continuous evaluation has widely been accepted in the philosophy of science (e.g., Lakatos & Musgrave, 1970). In one way or another, these validation boundaries play a role in controlling quality in scientific work at the level of the disciplinary fields and specialties (Mode 1).

Let us use the concept of validation boundary for exploring the potentially different mechanism of validation in Mode 2 research. One may expect that the quality of Mode 2 knowledge production can be validated independently of the Mode 1 tradition if the two control systems are different. However, the differentiation between the two forms of quality control can also be considered as a historical variable. When the two systems are differentiated, one is no longer able to rate them on a single scale. The validation boundary for Mode 1 (scientific excellence) can increasingly be independent from that for Mode 2 (that is, the relevance for users). One would then no longer be able to say which one is of lower or higher quality without specifying a system of reference.

For example, Mode 2 research does not always need to be subjected to the review process of traditional disciplinary publication. In this context, the concept of validation boundary is expanded to a boundary within which the knowledge is validated by a certain community (that is, a constitutive audience) and for a specific goal. The validation boundary of Mode 1 as explained above can then be considered a validation boundary in a narrow sense, validated only by the scientific community and with the sole purpose of stimulating "scientific excellence." The validation boundary of Mode 2, on the other hand, can be validated by the public or by users of the knowledge, for a specific goal, e.g. for the purpose of problem solving in the public sphere as explained in a following section.

Since one has initially few evaluation methods available for the transdisciplinary network activity of Mode 2 research, the quality control of Mode 2 is first modeled using the same validation boundaries which have been successful in Mode 1 research, that is, evaluation only by the results of publication. This arrangement, however, can lead to serious problems in the case of Mode 2 research. For example, Mode 2 researchers may individually be eager to produce publications in order to earn Mode 1 type of credit. Their achievements in terms of the numbers of publications may still function as keys to their future careers.

Note that these problems were also discussed as a central theme in the area of higher education in "inter- and transdisciplinary" studies during the 1970s, long before the discussion of this distinction between the Two Modes prevailed (e.g., CERI/OECD, 1972). Under the one-dimensional, publication-numbers-oriented evaluation systems, students who work in inter- and transdisciplinary subjects are likely to engage in submitting their papers to traditional, established journals rather than being involved in problem-solving. From the perspective of educational reform in inter- and transdisciplinary fields as well as from the perspective of the funding agencies of research, problem-solution can be the desired outcome of the efforts. The tendency to relate back to the "mother"-disciplines has often frustrated inter- and transdisciplinary research and education.

For example, Hayashi and Fujigaki (1999) compared the impact factors for journals in fields that can be considered typically Mode 1 (e.g., physics) with the impact factors for journals in fields of the Mode 2-type (e.g., artificial intelligence). The impact-factors for journals in Mode 1 fields were found to be higher than those for the Mode 2-type journals. These results might seem to suggest that Mode 1 research is oriented towards the peer review system and is therefore of higher quality, but we have to question the validity of this kind of one-dimensional quality judgement.

As mentioned above, the validation boundaries for Mode 1 and that for Mode 2 (user relevance) can grow into relative independence of one another, and one would thereafter no longer be able to say which one is of lower or higher quality. Fields of science are well known to differ in terms of the average impact factors of their respective journals. The one-dimensional evaluation system, which is oriented towards counting the numbers of publications, has erroneously produced the illusion of such a kind of one-dimensional quality judgement.

**Quality Control for Mode 2 Research**

One can study the value of Mode 2 research without using *a priori* the same criteria for quality control that are used for Mode 1 research. But one then has to clarify further the meaning of "value in use." To do this, one can first distinguish two major systems of reference: one for the market and one for problem-solving in the public sphere. In the first case one assumes that the market demand is representative of the users of knowledge and that the usefulness of scientific knowledge can thus be evaluated by the market. In the second case, one considers the "public" and not the market to be the main user of knowledge or the process of knowledge production. Some public issues may not be solvable in accordance with market forces. The usability of scientific knowledge and expertise, however, can be evaluated in terms of the ability to solve problems in the public sphere.

*Market force*
In general, one can consider the "market" as providing a scale of "value of use," which is different from scientific excellence. How is one able to use the "market rule" –that is, the capability of commercializing the products in a knowledge-based economy– as a standard for the evaluation of scientific research and development? Rip (1997) noted that the combination of scientifically excellent and socially relevant research may occur at the site of research. A successful combination of scientific excellence and applicational relevance can be considered as a resonance or a case of coincidence between two (analytically distinguishable) validation boundaries.

The relative degrees of combination and separation of standards for scientific excellence and user-orientation of the research depend on the organization of the tasks in the specific research fields (Whitley 1984). In some fields the two dimensions can easily be brought into harmony, especially when the market rule is highly applicable. Several examples are listed in Table 1.

A specific solution to the puzzle of combining scientific excellence with user-orientation can be achieved in research fields in which the products based on knowledge can be put on the market, for example, as patents or commodities. In fields to which the market rule cannot be applied, however, such as astronomy and the theory of elementary particles, one expects a separation between scientific excellence and user-orientation in research. In such a case, instrumentation can, for example, give rise to specific user-relations (Price 1984; Shinn 1997).

Table 1: Field differences in combinations of scientific excellence and in application relevance

| Fields | Combinations* | Usefulness** of Market-rule |
| --- | --- | --- |
| Advanced material | High | High |
| Chemistry | High | High |
| Biotechnology | High | High |
| Theory of elementary particles | No | Low |
| Astronomy | Little | Low |
| Health Science | ? | ? |

*Combinations between scientific excellence and user-oriented research within the knowledge production process
**Usefulness means applicability and infiltration of market rule towards outputs

Increasingly, governments tend to focus priority funding on the fields in which one can *combine* the two validation boundaries. Thus, the problem from the perspective of policy analysis is not the possibility of a coincidence between scientific excellence and (social and/or economic) relevance in some fields, but the fact that the mode theory causes governments to favor this specific coincidence in the case of priority funding. This tendency in government funding poses a problem for fields that intrinsically have little room for recombination but that can still contribute significantly to the solution of current problems.

*Problem-solving in the public sphere*

Users of knowledge may not be only the consumers on the market, but they can also be the people involved in problem-solving in public sphere. What is the validation boundary in problem-solving for public purposes? Examples of these kinds of problems are questions like: "How can one set standards for chemicals in industrial disposals in a way that will protect future environments?"; "How can one keep workers healthy under conditions of occupational stress?"; "How should one organize tax incentives to encourage patenting by the university sector?" Reports of research into these questions can be categorized as reports about Mode 2 research.

How can one organize quality control for this type of "contributions to problem solving"? It is likely to be quality control of a type different from that of Mode 1. How can one measure or otherwise evaluate the contribution to problem solving? The further analysis of the concept of "regulatory science" (Jasanoff 1990) provides us with a perspective on the evaluation of the contribution to problem solving.

In regulatory sciences, one can observe the operation of values other than publication-oriented evaluation and "pure" peer review. Jasanoff (1990) contrasts "regulatory science" with "research science." Regulatory science can be used for public problem-solving, e.g., for answering such questions as "How should standards for chemicals in industrial disposals be set in order to protect future environments?" Governmental agencies, citizens, and related stakeholders are then expected to be involved in the agenda-building and problem-solving processes. The contrasts between research science and regulatory science can be manifest in differences of products, standards and methods, review and validation processes, and time constraints.

In research science, products are publications of new and significant findings. Standards and methods are established by the relevant scientific communities. In regulatory science, products are studies of significant policy problems and assessments of scientific literatures. Standards and methods are sometimes established by regulation or by private professional organizations. Furthermore, in research science, review and validation is done through peer review by the experts in the field and through an open-ended process, without time limits. On the contrary, in regulatory science review and validation are done through expert judgements by scientific advisory committees and through criticism by interested parties under fixed policy timetables. Most importantly, the outputs of the regulatory sciences are controlled by validation boundaries in a broader sense, since a validation boundary in the knowledge will have to be acknowledged by several stakeholders with potentially different goals.

*Synthesis -- Linking Validation Boundaries*

How can one distinguish the epistemological status of expert judgement by scientific advisory committees and criticism by interested parties in terms of "validation boundaries" in the case of Mode 2 research? In regulatory science, or –more generally– in the case of problem solving for a public purpose, quality control and validation may be decided by citizens, advisory committees, and other interested parties. These groups of people, however, will entertain different opinions. We should thus consider how one can deal with the communication gaps between people who have different values and therefore impose different validation boundaries.

As explained above, the expanded validation boundary is defined as a boundary in which knowledge is validated by a certain community for a specific goal. How can the knowledge from different aspects be integrated for the purpose of problem solving? This integration would require the synthesis of several validation boundaries, since these problems cannot be solved within a single disciplines nor by a single and unique validation boundary.

Let us consider, for example, "consensus conferences" (Kobayashi and Wakamatsu, 1998). These meetings can be considered Mode 2 activities, since they provide the public with scientific information that serves as a basis for "problem solving for the public" (regulations on environmental issues, regulations on gene operation, and so on). But who is responsible for the quality of this information? How can this knowledge process be controlled and validated?

In our opinion, the output of a consensus conference is not expected to reflect a single perspective but can be considered as the mutual adjustment of a number of perspectives. For this adjustment of several perspectives, one can use a synthesis of several Mode 1 validation boundaries. However, for a synthesis one is in need of information from as many sources as possible and one has to devise alternatives that provide legitimization for a decision in the eyes of all the relevant audiences. While an "analysis" can be pursued on the basis of a specific validation criterion, "synthesis" requires classification and integration of several validation boundaries in order to address suggestions for future plans (Fujigaki, 1996). The "synthesis" should allow the different parties involved to move forward using their respective perspectives in partial consonance with the agreement.

Let us provide an example of the classification and integration of validation boundaries by elaborating on the issue of occupational stress. In this case, the research question that had to be solved was "How is one able to keep workers subjected to occupational stress healthy?" This question was raised by the Japanese Ministry of Labour (Fujigaki, 1997). To answer this question, the following procedures are used:

1. One first defines "stress conditions" and "not - stressed conditions" of work
2. One measures the stress conditions at work sites. (At the same time, one should increasingly standardize the measurement.)
3. One needs to establish criteria for permissible levels of occupational stress
4. One considers stress countermeasures in relation to the stress conditions measured at each work site.

During this process, however, one cannot apply analytical (supra-historical) measures to concrete situations: one has always to further specify the standards while developing the research locally, since the uniqueness of the situation requires a reflexive application of the measurement. Similarly, the counter-measures have to be tested carefully because one expects, given the complexities in the situation and its various contexts, unintended consequences.

For steps 1 and 2 of this procedure, the classification and integration of different validation boundaries are inevitable, since each measurement and definition has its own characteristics and tacit premises contingent on the conditions when the knowledge is constructed. These latter characteristics and premises can also be considered specific "validation boundaries" for the measurement.

Table 2 provides an example of such a summation of validation boundaries. It classifies the characteristics of different measurements related to the study of stress (Tokyo Declaration on Work-Related Stress and Health, 1998). This table also summarizes the characteristics of measurement utilized for the occupational stress issues to be solved. Using the table, one is thus able to predict the strengths, weaknesses, and application-ranges of each measurement. More concretely, it exhibits application targets (for individuals and for collectives), sensitivity in the time dimension, applicable disease phases, and sensitivity for factor finding in each specific type of measurement. The table reveals the condition contingent with the measurement. Thus, the table guides the decision process about suitable measurements instruments or counter-measurement instruments for each work-site with its specific characteristics.

Table 3: Example table summarizing the characteristics of measurements

---

| Measurement | Sensitive-time-dimension | Suitable-disease-phase | Suitable-task |
|---|---|---|---|
| Psychological method | | | |
| NIOSH, JCQ | week/month | cumulative | general |
| ⋮ | | | |
| Physiological method | | | |
| Heat-rate-variability | second/min | instantaneous | pilot-workload |
| ⋮ | cumulative | patients | |
| Biochemical method | hours/days | cumulative | workers with cumulative fatigue |
| ⋮ | | | |

---

This type of table can also be used when one makes guidelines for consensus meetings with public audiences. (For example, guidelines for environmental pollution, and guidelines for accountability in medical treatment.) It can function as a checklist of different validation boundaries.

In the case of transdisciplinary knowledge production, it is important to find optimum solutions for given boundary conditions within a limited time. Moreover, if we want to find an optimum solution, we need to gather as much relevant information as possible and to bring all alternatives on the table. Such a summarization of the validation boundaries of the related fields (including measurement) is illustrated in Table 2.

A very similar process can be observed in the case of research questions that address public problem-solving in the case of environmental regulations. In the case of air or water pollution, for example, one should (1) define the air/water pollution at each observation point, (2) measure the pollution by using standardized measurement, (3) establish criteria for maximum permissible levels of chemical compounds, and (4) consider environmental countermeasures for each country/company and establish international agreement. For the steps 1 and 2, validation boundaries of related fields including different tacit premises should be summarized as in Table 2.[3]

Thus, Mode 2 quality control is not organized as a final and optimal solution. It is a process of summarizing different validation boundaries, with the goal of making proposals, future plans, and more definitive decisions. Furthermore, the validation boundaries for these activities (making proposals, future plans, and final decisions) are constructed on grounds different from the criteria on which validation boundaries and quality control systems for Mode 1 research are constructed. The expanded validation boundaries for Mode 2 can be considered as a further integration of the validation boundaries in the narrow sense of Mode 1. The validation boundaries for Mode 2 constitute a next-order codification of the validation boundaries with reference to a transdisciplinary context.

**Institutional boundaries and validation boundaries**

These several meanings of "quality control in Mode 2" are inevitably based on discussions and negotiations at the interfaces of institutional actors like academia, industry, and government, and also at the interface of their interactions with "the public" at large (e.g., the media). The institutional boundaries and the validation boundaries for quality control, however, are not coupled structurally, but operationally.[4]

The different boundary formations --which operate upon each other-- are interfaced by culturally different (e.g., national) systems of innovation. The Triple Helix model (Etzkowitz and Leydesdorff, 1997) can be used to address boundary formation and spanning mechanisms between three major sectors; the university, industry, and government sectors. These sectors are differently shaped both culturally and historically. Furthermore, the cultural and historical differences affect the role of academia and the configuration of Mode 2 research in each nation state (Hirasawa *et al.*, 1998; Shinn 1998).

In the case of Japan, for example, the three sectors have remained largely independent and there has been little exchange between them –especially during the 1970s; that is, in the aftermath of the students' movement of the 1960s (OECD, 1977). University research, governmental institutes, and industry were said to be specialized for "basic science," "applied science," and "development," respectively. The economic prosperity of the 1980s was driven not by basic science in the universities, but by the knowledge creation process endogenous to the industries (Nonaka and Takeuchi, 1995).

This relative isolation of academics from the processes of social production and distribution is currently changing because of the global transition towards a knowledge-based economy and a corresponding research policy (Japanese University Council, 1998; Fujigaki and Nagata, 1998). But because boundaries of the university sector have become so thick in Japan during the past thirty years, one feels nowadays a strong need to promote mobility between sectors (23rd Recommendation by S&T Council, and law for promoting research exchange). Furthermore, some criticisms of Japanese academics have also been raised with respect to their inactive attitudes related to problem solving in public sphere (Yonemoto, 1999). These attitudes have been fostered by the "pure science" ideology prevailing in academia.

Thus, the thickness of the institutional boundaries affects the activity of Mode 2 research at the interface of the three sectors by isolating the validation boundary of Mode 1. As Gibbons *et al*. (1994) pointed out, the reflexive activities of researchers, which are involved in Mode 2 research, can have an effect on the validation boundary of Mode 1. Here, however, we see that conversely, an inactive attitude towards Mode 2 research may also have an effect on the solidity and isolation of the Mode 1 boundaries.

In the Netherlands, as a counter-example, during the 1950s, one had a system of thick boundaries between the "pillars" of society based on the various religions. In the 1960s, these pillars lost their legitimacy as an organizational principle, but they have provided the system with an intermediary layer for continuous negotiations. This also has had an effect on the discussion of the future of the universities. The various social partners feel free to raise issues, and there are channels for converting resolutions into compromises (Rip and Van der Meulen, 1996). More recently, networking at the level of the European Union has added another layer to this system, facilitating the transition of Mode 1 and Mode 2 research but in contrast to the way this transition is facilitated in Japan, which has remained very much a national system (Leydesdorff, 2000)

**The Future of the Academe**

The shifts from an institutional frame of reference to one focused on the dynamics of communications has enabled us to clarify the integration and differentiation of different forms of knowledge production, as well as to deconstruct one-dimensional quality judgements. For example, the synthetic integration of different validation boundaries in the case of stress research showed how the emerging validation boundaries are able to operate as next-order codification. Given this view of validation boundaries, what can one expect with respect to the future of academic research? What is a future outcome of combining and resolving conflicts among validation boundaries in the quality control of research?

During past decades, ceaseless questions about the "usefulness" of university research have stimulated university researchers to do both Mode 2 and Mode 1 types of research. However, these two types of research have different validation boundaries. Universities are, on one hand, expected to focus on the activities of classical disciplinary work and, on the other hand, to function as the "nodes" for networking in Mode 2 research.

For the former function, the classical "critical functions" of academics —that is, critical attitudes towards real-world problems in terms of social responsiveness as choices of individual academics—can be invoked. These critical functions also *include* the reexamination of the validation boundaries of Mode 1. Thus, these "enlightenment" values do not operate necessary in a conservative way.[5]

The reflexive activities of research, which are involved in Mode 2 research, can also have an effect on the validation boundary of Mode 1 research. Fuller (1999b), for example, objected to the suggestion implicit in the "Two Modes model" that academics would have to choose between disciplinary closure and openness to non-academic concerns. In our opinion, one is able to analyze disciplinary "openness" and closure to public purposes in terms of the interactions between validation boundaries in the narrow sense and the expanded sense.

Furthermore, validation boundaries at the level of the communication system interact with institutional boundaries. As mentioned above, the isolated institutional boundaries in Japan have had an effect on the solidity and isolation of the Mode 1 boundaries, and the inactive attitudes toward Mode 2 research have also been reinforced. The Japanese universities are now facing the prospect of changing from national universities to independent administrative agencies (or cooperation) (Basic Law on Administrative Reforms, 1998). This reform of institutional boundaries is intended to change the differentiation of knowledge production at the level of the respective "communication" systems.

In our opinion, the critical functions of the academy should be considered structurally as missions of research as well as in terms of choices of individual academics. The university's unique comparative advantage is that through the passage of student generations, it combines continuity with change, organizational memory and research memory with new individuals and new ideas. These retention and reproduction mechanisms are evolutionarily different from the attitudes irreflexively keeping and reproducing the Mode 1 validation boundaries in scientific journals. The accumulation of research memory for problem-solving for public purposes continuously constructs new validation boundaries at the level of the society, and these new boundaries have to be transmitted to the next generation of students. This restructuring has a reflexive and potentially critical function.

Nowadays, the universities are seeking a way to play a role in problem-solving for public purposes. They are also competing with other recently proposed contenders for knowledge leadership, such as consulting firms, which may also serve for solving public problems. Such firms, however, lack the organizational ability needed to pursue a cumulative research program as a matter of course (Etzkowitz & Leydesdorff, forthcoming). The university has become salient in a knowledge-based economy because of its potential to serve at the junction of higher education, research, and economic development. University research and higher education can act as "catalysts" for each other in the processes of translating different forms of code.

Students are also potential inventors. They represent a dynamic flow-through of "human capital" in academic research groups, and in this way they contrast with the more static industrial laboratories and research institutes. The university provides a "laboratory" of knowledge-intensive development and it is at the same time the main reproductive function of this system. The university can be expected to remain the core institution of the knowledge sector as long as it retains its original educational mission (Etzkowitz, Webster, Gebhardt, and Terra, forthcoming).

Because of this structural position between institutional reproduction and academic control, the university as an institution can remain "critical" to the system at the social level (Godin & Gingras, forthcoming). "Critical" here has a meaning at the level of this social system different from its meaning for the individual. It is more related to the original meaning of "decisive." While the traditional critical function of the university has emphasized the role of "individual" choices, we propose to consider the structural functions of universities in terms of its functions for the social communication. Both individual "criticality" and structural "criticality" may nowadays be required at the university in order to fulfill the envisaged combination of functions.

For example, while "interdisciplinarity" was often defined in terms of finding a common denominator or a normative orientation, the understanding of transdisciplinary communication problems in terms of reflexive "translations" now sets a different agenda for educational reform (Tobias *et al.* 1995; Leydesdorff & De Klerk 1998). Similarly, the gradual replacement of the laboratory model of innovation by a "desktop" model challenges the research agendas. This change of the innovation process itself asks also for further reflections about the policy instruments of the previous period (Kaghan and Barrett 1997). Academia will have a say in developing the instruments needed for improving and controlling quality.

Questions are raised on how the university can commit intellectually to the surrounding society as a mark of its social responsiveness and on how the university can be reconstructed institutionally in order to function as the junction of higher education, research, and economic development. Beyond the restriction to the traditional validation boundaries (Mode 1), the development of validation boundaries and quality control commits research and education to the larger society, while these yardsticks can at the same time be used for the further development of higher education and university research.

*Acknowledgements*

The authors wish to acknowledge partial funding by the SOJIS project (Special Co-ordinating Fund by STA, Japan) and by the program for Targeted Social-Economic Research of the European Commission, project nr. SOE1-DT97-1060 ("The Self-Organization of the European Information Society").

Lundvall, Bengt-Åke (1988). Innovation as an interactive process: from user-producer interaction to the national system of innovation, in G. Dosi, C. Freeman, R. Nelson, G. Silverberg, and L. Soete (eds) *Technical Change and Economic Theory,* London: Pinter, pp. 349-369.

Maturana, H. (1978) Biology of Language: The Epistemology of Reality in G. A. Miller and E. Lenneberg (eds) *Psychology and Biology of Language and Thought. Essays in Honor of Eric Lenneberg*, New York: Academic Press, pp. 27-63.

Maturana, H. R., & Varela, F. J. (1980). *Autopoiesis and Cognition: The Realization of the Living*. Dordrecht.: Reidel.

McKelvey, M. (1997) Emerging Environments in Biotechnology in H. Etkzowitz & L. Leydesdorff, *Universities and the Global Knowledge Economy: A Triple Helix of University-Industry-Government Relations*, London: Cassells Academic, pp. 60-70.

Nonaka, I. and Takeychi, H. (1995) *The Knowledge Creating Company: How Japanese Companies Create the Dynamics of Innovation.* Oxford: Oxford University Press.

Nowontny, H. (1990) Actor-networks versus Science as Self-Organizing System: A Comparative View of two Constructivist Approach, *Sociology of the Sciences*, Vol.14, 223-239.

OECD (1977) *Social Science Policy: Japan.* Paris.

Centre for Educational Research and Innovation CERI (1972). *Interdisplinarity*. Paris: OECD.

Parker, E. N. (1997) EOS (Newsletter of American Association of Geophysics), 79, No.37,391

Price, D. S. (1984) The science/technology relationship, the craft of experimental science, and policy for the improvement of high technology innovation, *Research Policy* 13: 3-20.

Rip, A. (1997) A Cognitive Approach to Relevance of Science, Social Science Information, 36(4), 615-640.

Rip, A. and van der Meulen, B.J.R. (1996) The Post Modern Research System, *Science and Public Policy*, Vol.23, No.6, 343-352.
20

**Notes**

---

[1] The concept of validation boundaries was derived by applying autopoiesis theory to the production of scientific knowledge (Fujigaki, 1998). A journal and a scientific paper can respectively be

considered as a unit and a component of a scientific communication system. The journal system is then defined as a chain of publications, in which each paper is a component of the system –each component (paper) leads to the production of the next one (the next paper)— and this production process operates continuously. Nowotny (1990) noted that the autopoietic systems theory can be considered one of the two dominant constructivists' approaches. The other is actor-network theory (Callon *et al*.; 1986). Autopoietic systems theory can provide a bridge between the site of knowledge production (scientific paper) and the site of theory making in science studies.

[2] "Peer review" has been identified by Fuller (1999a) as:
1. a mechanism for certifying research activity as knowledge and for crediting researchers with having produced knowledge (e.g., journal editing policies)
2. a mechanism for protecting the knowledge base from error contamination and the public from the application of unsound research (e.g., state regulatory bodies, ethics review panels)
3. a mechanism for the efficient and equitable allocation of the scarce resource available and needed conducting research (e.g., funding agencies).

[3] This kind of table can also be useful for settlement of cross-disciplinary conflicts. Actually, Table 2 served also as a bridging mechanism in a cross-disciplinary conflict between the researchers from different disciplines.

[4] One is able to distinguish between "structural coupling," "operational coupling," and "loose coupling" (Leydesdorff 1994). Two systems can be considered "structurally coupled" if the operation of one system disturbs the operation of the other system, as in co-variation or co-evolution (Maturana 1978). A structural coupling is operationally closed (Luhmann 1984). "Operational coupling" means that the two systems are interfaced, for example, by a common network. At a larger distance, systems are only loosely coupled (Simon 1969).

[5] Three dimensions can be distinguished in discussions about the future of academic research: (1) ethical norms at the level of the profession, (2) knowledge legitimization and authorization, and (3) usefulness for social need (e.g., market force and public problem-solving). In this paper, we focused on the second and the third dimension, while the first two have been more central to the discussion about traditional roles for academic research. For example, Gieryn's (1994) discussion of "boundary-work" relates the norms of the profession (first dimension) to knowledge legitimization and authorization (second dimension). Using the three dimensions, one may be able to describe the social relations of academic research as a complex dynamics.